# CMS Token Transition

*Brian* Bockelman[a1], *Rahul* Chauhan[b], *Diego* Ciangottini[c], *Dave* Dykstra[d], *Edita* Kizinevic[b], *Stephan* Lammel[d2], *Marco* Mascheroni[e], *Sarun* Nuntaviriyakul[b], *Panos* Paparrigopoulos[b], *Alan* Malta Rodrigues[f], *Chan-anun* Rungphitakchai[g], *Eric* Vaandering[d], *Vaiva* Zokaite[b]

[a]Morgridge Institute for Research, 330 N Orchard Street, Madison WI, USA
[b]CERN, European Organization for Nuclear Research, Espl. des Particules 1, 1217 Geneve, Switzerland
[c]INFN, Perugia, Umbria, Italy
[d]Fermilab National Accelerator Laboratory, Wilson Road, Batavia, IL 60510, USA
[e]University of California at San Diego, San Diego, CA, USA
[f]University of Notre Dame, Notre Dame, IN, USA
[g]University of Wisconsin, Madison, WI, USA

**Abstract.** Within the LHC community, a momentous transition has been occurring in authorization. For nearly 20 years, services within the Worldwide LHC Computing Grid (WLCG) have been authorized based on mapping an identity, derived from an X.509 credential, or a group/role, derived from a VOMS extension issued by the experiment. A fundamental shift is occurring to capabilities: the credential, a bearer token, asserts the authorizations of the bearer, not the identity. By the HL-LHC era, the CMS experiment plans for the transition to tokens, based on the WLCG Common JSON Web Token profile, to be complete. Services in the technology architecture include the INDIGO Identity and Access Management server to issue tokens; a HashiCorp Vault server to store and refresh access tokens for users and jobs; a managed token bastion server to push credentials to the HTCondor CredMon service; and HTCondor to maintain valid tokens in long-running batch jobs. We will describe the transition plans of the experiment, current status, configuration of the central authorization server, lessons learned in commissioning token-based access with sites, and operational experience using tokens for both job submissions and file transfers.

[1] Corresponding author: bbockelman@morgridge.org
[2] Corresponding author: lammel@cern.ch

# 1 Introduction

The Compact Muon Solenoid (CMS) collaboration operates one of the general purpose detectors at the Large Hadron Collider (LHC) at CERN in Geneva, Switzerland. During proton-proton operation the experiment expects to record between 140 to 400 PBytes [1] per year during the high-luminosity (HL-LHC) collider operation. To reconstruct physics quantities from the detector data, a vast amount of computing is required. Since the beginning of the LHC era in the early 2000s, high-energy physics (HEP) experiments, like CMS, have used grid computing. Computing authorization on the Worldwide LHC Computing Grid (WLCG) is based on X.509 certificates [2] at both client and services sides. Use of X.509 for user identification did not receive much acceptance outside HEP, and support for proxy certificates to prove experiment or group membership [3] is declining. Instead, web services converged on bearer tokens acquired via OAuth 2.0 [4] workflows to authorize service access.

Tokens bring a fundamental change to grid authorization: with a personal X.509 certificate the user proved her/his identity (and experiment/group membership) on the grid and the service decided to permit or reject an operation. Tokens are issued by the experiment and authorize a specific operation at a service independent of the bearer.

CMS embraced this change in authorization philosophy, seeing the opportunity to benefit from a more secure computing infrastructure. The migration from X.509 to tokens is ongoing and expected to be complete well before the start of HL-LHC in 2030.

# 2 CMS Authorization Needs

Token-based authorization must be able to fully substitute authorization currently handled via user X.509 certificates. We expect the use of X.509 certificates for host/service proof of identity to continue as currently even after full token migration. The token infrastructure needs to be robust. Outages of the CMS computing grid should remain well below 1% during a year. The robustness of the token infrastructure then determines the lifetime of access tokens. For an access token lifetime of one week, equivalent to the current X.509 user proxy lifetime, the token infrastructure outage should be less than a few days per year. For an access token lifetime of an hour, the aim, the token infrastructure needs to be available 99.7% of the time.

The three main activities in CMS that critically depend on authorization and are expected to require the most authorization include 1) data transfers, i.e., distributing detector, reconstruction, and simulation data for processing and analysis within the CMS computing grid, 2) accessing the Compute Elements to start jobs on worker nodes, and 3) accessing storage from the worker nodes within the CMS computing grid.

CMS uses Rucio and FTS for data transfers. In 2024 CMS transferred on average 900,000 files per day with peaks of 2.9M files. Those files belonged to 332K datasets. Authorization for a transfer requires three access tokens, one to authorize with FTS, one to authorize file reading at the source, and one to authorize file writing at the destination site. In the current Rucio/FTS token implementation FTS acquires file read and write refresh and access tokens itself, for a total of seven tokens per transfer. For file-based tokens, this equates to over 6M tokens per day, with an average frequency of around 70 tokens per second (70 Hz). For dataset-based write tokens and read tokens, the creation rate is 40K tokens per day for a 1-week lifetime, which corresponds to an average frequency of approximately 0.46 Hz. During steady running with transfer queues in FTS of less than an hour, files in the same transfer request will be handled by the same set of tokens. For tokens with a 1-hour lifetime, the creation rate increases to 50K tokens per day, translating to an average frequency of around 0.58 Hz.

During HL-LHC, CMS expects to record about 400 PB of collision data per year, with about twice as much derived data. Transferring these data will require about 350,000 tokens per day or 4 Hz of one-hour-lifetime access tokens and 1.5 Hz of week-long refresh tokens.

The next area where tokens play a pivotal role is job submission, which allows running reconstruction, simulation, and analysis jobs via a pilot-based workload management system. Pilot jobs are submitted to Compute Elements to reserve resources on worker nodes, which then allows the scheduling of the reconstruction, simulation, or analysis jobs. To access the Compute Elements, X.509 certificates and tokens are currently being used. Assuming a moderate increase in Compute Elements during HL-LHC, submitting the pilots would then require about 1,500 tokens per hour, or 0.4 Hz of one-hour lifetime access tokens.

Lastly, storage access on the worker node has two modes of operation: reading data from storage and writing data to storage (stage-out). During 2024, CMS used between half a million to a million worker node cores at any given time. About two third of the cores were used by data reconstruction and simulation jobs utilizing 2, 4, or 8 CPU cores. The other third by single-core analysis jobs. Both types of workflows use multiple jobs to process a dataset in parallel.

During HL-LHC, CMS expects to use several million CPU cores with on average jobs utilizing more cores than currently. The number of jobs to process a dataset will likely increase too, as datasets are larger and more complex to process. The Workload Management System is planned to use one token for each workflow, with both read and write scopes enabled for the relevant datasets. Depending on the composition of workflows executing at a specific time, storage access from grid worker nodes will require between 35,000 and 200,000 access tokens per hour or 10 to 55 Hz of one-hour-lifetime access tokens.

In summary, the current authorization plan of CMS requires a moderate rate of access and refresh tokens from the token issuer at tens of Hertz during HL-LHC operations in the 2030s.

## 3 CMS Token Infrastructure

The CMS token-based authorization approach consists of 1) an INDIGO Identity and Access Management (IAM) service to issue tokens, 2) storage and compute services at sites permitting operations according to the authorization in the token, 3) a HashiCorp Vault (HCVault) service to store and refresh access tokens for users, 4) a managed token bastion to update the HTCondor CredMon service with vault access credentials, 5) the HTCondor Credd service to supply the potentially long-running jobs with valid short-lifetime tokens, and 6) utilities to shield users from token handling/management.

Fig 1 shows the envisioned token flow. The HashiCorp Vault acquires access tokens from the IAM service via refresh tokens. The managed token bastion fetches vault and access tokens from the HCVault and updates the CredMon service on the HTCondor schedd machines with vault tokens. CredMon maintains valid access tokens for the Credd service acquired from the HCVault. The Credd service supplies CMS jobs running on grid worker nodes with storage access tokens,

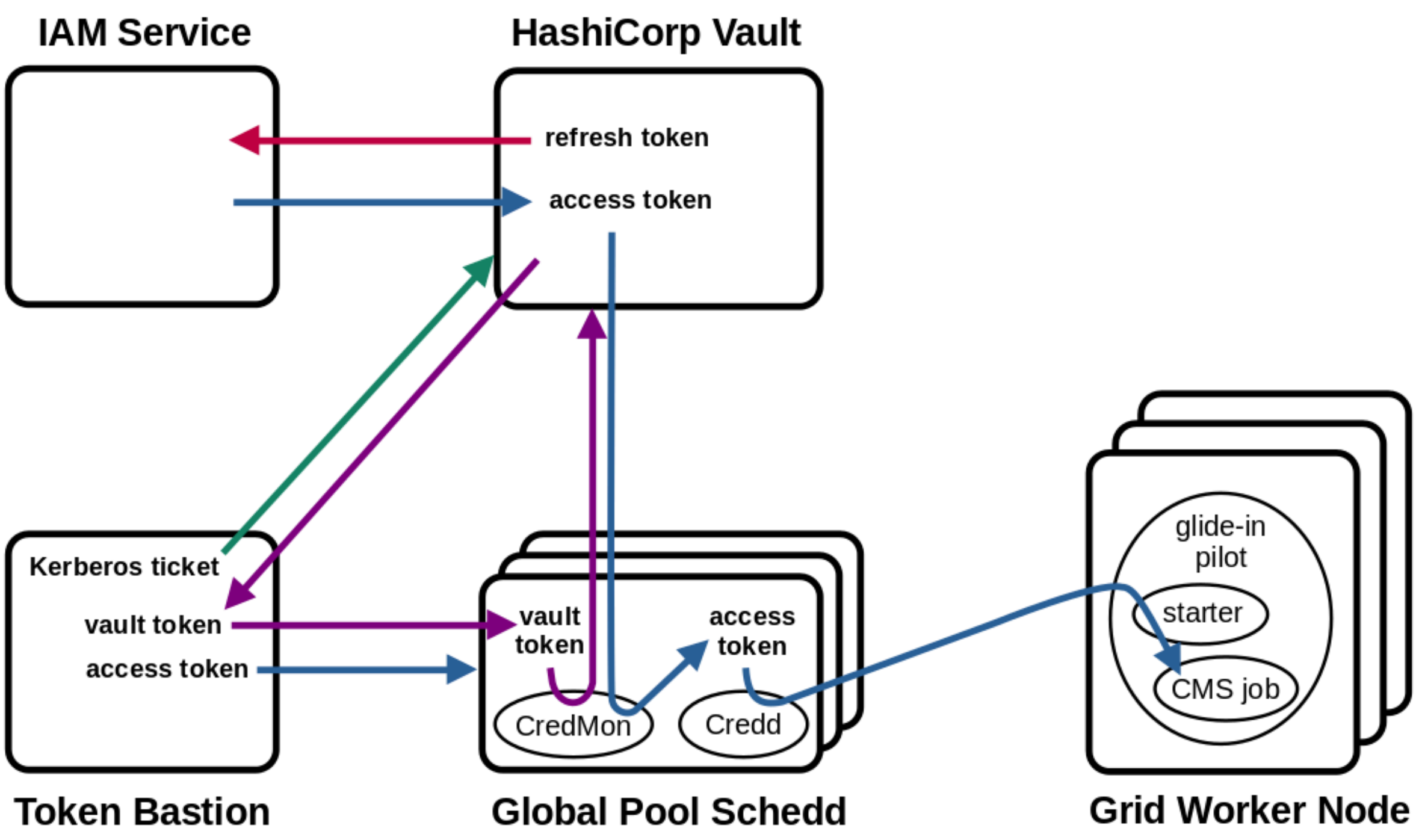

**Fig.1.** Components of the planned CMS token infrastructure with credentials used for authorization and token flow from the IAM service via HCVault, CredMon, and Credd to the reconstruction, simulation, and analysis jobs on grid worker nodes.

### 3.1 INDIGO Identity and Access Management Service

In 2017, WLCG created a working group to investigate the evolving authentication and authorization landscape and assess how WLCG could benefit from advancements there. A common JSON Web Token schema and profile was developed [5] to facilitate interoperability beyond WLCG. In 2019, the INDIGO Identity and Access Management Service [6] (IAM) was selected as the token issuer for WLCG. The software is developed and maintained by INFN.

Instances for WLCG are deployed on the CERN IT infrastructure where a dedicated team is maintaining an IAM instance for each of the LHC experiments.

CMS integrates IAM with its existing authentication and authorization infrastructure. The CERN e-group system is used to maintain computing and group membership information, which is imported into the Computing Resource Information Catalog (CRIC) [7]. CRIC serves as a central repository that synchronizes computing roles and user account information by querying CERN's Human Resource (HR) database and CERN e-groups via LDAP.

A synchronization mechanism ensures that IAM receives the necessary user and group information from CRIC. A periodic synchronization script runs every three hours, automatically updating IAM with new users, updated group memberships, and certificate details. The script extracts relevant data from CRIC, such as user CERN IDs, associated certificates, and group assignments, and pushes this information into IAM. This automation ensures that new CMS members are seamlessly integrated into IAM without requiring manual intervention.

IAM authorization policies are centrally defined within IAM. Each IAM group is linked to a specific scope policy that dictates the permissions granted to its members. While the policy definitions reside in IAM, group membership is managed in CRIC. CERN e-groups are synchronized into CRIC groups, and specific CRIC groups are then synchronized with IAM. This structured synchronization ensures that users in the relevant CRIC groups automatically receive the corresponding IAM policies without requiring direct intervention in IAM.

By maintaining IAM group memberships through CRIC, CMS ensures that authorization policies remain consistent and manageable across its distributed infrastructure. The IAM scope policies dictate access levels and permissions associated with different roles, such as job submission, data access, and storage management. CRIC serves as the authoritative source for group assignments, and any modifications to these groups are propagated to IAM via the synchronization process. This structured approach allows CMS to maintain fine-grained control over token-based access while avoiding direct manual configuration in IAM.

Users authenticate to IAM via the CERN Single Sign-On (SSO) service [8], ensuring a seamless and secure login process. This authentication mechanism allows users to verify their account status, request tokens, and authorize clients without additional credential management overhead. By integrating IAM tightly with CRIC, CMS has established a robust and automated system for managing user identities, group memberships, and token-based access policies.

### 3.2 Storage and Compute Services

CMS uses two main compute service technologies to access the Compute Elements (CE) for pilot job submission: ARC-CE [9] from the NorduGrid community and HTCondor-CE [10] from the HTCondor team.

When reconstruction, simulation, or analysis jobs are submitted, central services, called factories, send pilot jobs to the Compute Elements using the HTCondor grid universe and local scheduler daemons. An ARC proprietary client provided by HTCondor, and an HTCondor client are used on the factories to respectively access ARC-CEs and HTCondor-CEs using tokens. These tokens are retrieved from IAM using dedicated clients registered on the IAM servers. On the Compute Element side, the client ID of the token is checked against a list of pre-staged client ID values. The validity of the token is currently one week.

Upgrade and configuration campaigns for CE token support started in 2022. Submission to all HTCondor-CEs is based on IAM-issued tokens since the end of 2023. The migration has been delayed by an ARC-CE submit client issue, and factories still use X.509 certificates for submission to ARC-CEs as of this writing. In preparation for the transition, integration tests using tokens on ARC-CEs have been successfully done.

For storage transfer and access, CMS uses the *xroot* [11] and *WebDAV* [12] protocols. The former is a protocol developed by the HEP community for efficient wide-area network data access and the latter is an extension of the *https* protocol for data access/transfer. Support for token based authorization was added to the various storage implementations over the last years. CMS set up detailed tests within the WLCG Experiment Test Framework (ETF) [13] before the commissioning campaign. ETF runs a test against the configured site storages and provides results and failure details including commands to reproduce an issue. The ETF tests eliminated the needs for most sites to acquire tokens themselves. During 2023 and 2024 storage at sites was upgraded and configured. Except for EOS [14] where a token access issue delayed the commissioning campaign, all site storages accept IAM-issued tokens for access authorization.

| State | Service | Icons | Status detail | Age | Checked | Perf-O-Meter |
|---|---|---|---|---|---|---|
| OK | org.cms.SE-WebDAV-1connection | | OK - Endpoint reachable on all addresses | 17 h | 14 m | |
| OK | org.cms.SE-WebDAV-2ssl | | OK - SSL access to endpoint | 17 h | 14 m | |
| OK | org.cms.SE-WebDAV-3crt_extension | | OK - WebDAV protocol extension supported at endpoint | 17 h | 14 m | |
| OK | org.cms.SE-WebDAV-4crt-read | | OK - file read access test successful | 17 h | 14 m | |
| OK | org.cms.SE-WebDAV-6crt-access | | OK - file open-access test successful | 17 h | 14 m | |
| OK | org.cms.SE-WebDAV-7crt-write | | OK - file write access test successful | 17 h | 14 m | |
| OK | org.cms.SE-WebDAV-8crt-directory | | OK - directory operation test successful | 17 h | 14 m | |
| OK | org.cms.SE-WebDAV-10macaroon | | OK - Macaroon support test successful | 17 h | 14 m | |
| OK | org.cms.SE-WebDAV-14tkn-read | | OK - file read test successful | 17 h | 14 m | |
| OK | org.cms.SE-WebDAV-16tkn-access | | OK - file access restricted | 17 h | 14 m | |
| OK | org.cms.SE-WebDAV-17tkn-write | | OK - file write test successful | 17 h | 14 m | |
| OK | org.cms.SE-WebDAV-18tkn-directory | | OK - token directory operation test successful | 17 h | 14 m | |
| OK | org.cms.SE-WebDAV-99summary | | OK - WebDAV probe successful | 17 h | 14 m | |
| OK | org.cms.SE-XRootD-1connection | | OK - Endpoint reachable on all addresses | 17 h | 217 s | |
| OK | org.cms.SE-XRootD-3version | | OK - version test successful | 17 h | 217 s | |
| OK | org.cms.SE-XRootD-4crt-read | | OK - file read test successful | 17 h | 217 s | |
| OK | org.cms.SE-XRootD-5crt-contain | | OK - foreign files inaccessible | 17 h | 216 s | |
| OK | org.cms.SE-XRootD-6crt-access | | OK - file access restricted | 17 h | 216 s | |
| OK | org.cms.SE-XRootD-7crt-write | | OK - file write test successful | 17 h | 216 s | |
| OK | org.cms.SE-XRootD-8crt-directory | | OK - directory operation test successful | 17 h | 216 s | |
| OK | org.cms.SE-XRootD-9federation | | OK - files reachable via federation | 17 h | 216 s | |
| CRIT | org.cms.SE-XRootD-14tkn-read | | CRITICAL - file stat error | 17 h | 216 s | |
| UNKN | org.cms.SE-XRootD-15tkn-contain | | UNKNOWN - prerequisite test failed | 17 h | 216 s | |
| UNKN | org.cms.SE-XRootD-16tkn-access | | UNKNOWN - prerequisite test failed | 17 h | 216 s | |
| CRIT | org.cms.SE-XRootD-17tkn-write | | CRITICAL - GFAL2 file open/write/stat failure | 17 h | 216 s | |
| UNKN | org.cms.SE-XRootD-18tkn-directory | | UNKNOWN - prerequisite test failed | 17 h | 216 s | |
| OK | org.cms.SE-XRootD-99summary | | OK - XRootD probe successful | 17 h | 218 s | |

**Fig. 2.** ETF tests result of the CERN WebDAV and XrootD EOS endpoints. The XRootD token tests show the remaining access issue.

Data management in CMS is based on Rucio [15] with transfers submitted to the CERN File Transfer Service (FTS) [16]. Token support was added to both for the WLCG data challenge in early 2024. To transfer files between site storage, Rucio acquires a token to read from the source and write to the destination storage and submits a transfer request to FTS. FTS exchanges the two access tokens for refresh tokens and maintains those until the transfer is being scheduled. At that time, access tokens are derived and the transfer performed. CMS switched to token-based transfers in autumn of 2024 after a security update to IAM, for source and destination storage types other than EOS.

### 3.3 HashiCorp Vault Service

Compared to X.509 certificates, tokens are short lived and usually acquired via OAuth 2.0 [4] workflows. Shielding the user from token management is one of the CMS goals. A HashiCorp vault is used to store longer lifetime refresh tokens securely and acquire access tokens for the user as needed.

CERN IT is already using a HashiCorp vault for secret management and agreed to also run a vault service based on enhancements made by Fermilab for its experiments [17], along with CMS-specific scope configuration. The enhancements from Fermilab are distributed in a package called htvault-config [18] so the service is often referred to as HTVault.

The HTVault service is configured to initially authenticate with CMS IAM using an OAuth 2.0 standard extension protocol called OpenID-Connect (OIDC), through a web browser. That process stores a long-lived renewable refresh token in HTVault and returns a vault token and bearer token to the user. That vault token lasts for an intermediate amount of time (1 week), but renewed access to the longer term (4 weeks, renewable) refresh token stored in HTvault can be obtained through Kerberos, which is automatically available to command line tools by default for CERN users.

The command line client that communicates with HTVault is called htgettoken [19]. It manages the different types of authentication and automates the flows with HTVault.

### 3.4 Managed Token Bastion

Processing and reconstruction jobs of the CMS experiment run unattended 24 hours a day. For those jobs (and also user jobs in the initial phase), a separate service is needed for long-lived credentials; this service ensures credentials are renewed and minimizes the servers where such credentials are kept. A managed token bastion holds Kerberos credentials and pushes valid Vault access tokens to the “entry points” of the batch system, where jobs are submitted. CMS uses an HTCondor pool for its batch system, see above. The entry points for jobs are the schedd machines, that are not directly accessible to users but to the agents of the production system [20] and the task workers of the CRAB client-server tool [21].

The managed token bastion is implemented in python [22]. It works based on a configuration and maintains a valid vault token for each account and token scopes it manages at the relevant schedd machines of the pool. Updates at the schedd machines are done via access token authorization.

As of the writing of this paper, the managed token bastion is currently being deployed and configured.

### 3.5 HTCondor CredMon and Credd Services

Part of the Fermilab HTVault work was also an integration of the HashiCorp vault service and htgettoken with HTCondor. A condor_credmon_vault daemon was developed that maintains valid access tokens at the schedd, along with a condor_vault_storer callout from condor_submit. condor_vault_storer invokes htgettoken and stores vault tokens into condor_credmon_vault which acquires access tokens from HTVault as needed. However, for the initial CMS use cases the managed token bastion will store the vault tokens into condor_credmon_vault instead, since credentials for individual researchers are not yet needed. They will be needed for other use cases in the future (see next section).

The condor_credmon_vault daemon is essentially a vault-specific plugin to the standard HTCondor CredMon service, and CredMon works in tandem with the HTCondor Credd service. The Credd service transports the short-lived access tokens to the long-running jobs on the worker nodes around the globe to maintain valid access tokens inside the job, keeping the access tokens fresh in the jobs. It is a similar process to the X.509 proxy certificates, but given that access tokens have a shorter lifetime, updates from the Credd to the grid job happen more often.

We have exercised the full chain of tokens acquired from the vault via CredMon and Credd to user jobs. We are now waiting for the managed token bastion to move toward a production setup for first job types.

### 3.6 Utilities for Users Token Handling/ Management

Our current implementation has not yet reached the stage that users need to have tokens for the commands they use. We expect this to come as we "tokenize" CMS services. Work on this will start in the second half of 2025 after the CMS centralized web services accepts tokens. The “CMS web” service provides the authorization front-end for many client-server tools of CMS.

## 4 Transition Plans and Current Status

We need to resolve the remaining issues with ARC-CE job submission and the access issue on EOS; this is ongoing. Upgrades to the stage-out procedure at the end of grid jobs to handle tokens, i.e. to try an X.509 based transfer in case of token-based transfer failure, has started. Once those two are complete, grid jobs can be supplied with storage access tokens. Running jobs with both X.509 and token credentials will allow us to verify the robustness of the token infrastructure (while having a fallback to the current X.509 approach) and adjust components as needed.

Token access on our central web service requires many experiment-specific scopes. We discussed the best scope organization with the IAM team, to make sure policy rules can be written and are easily understandable. We will prototype and compare different organizations at the upcoming IAM hackathon in February of 2025. By the end of 2025 we expect to have the first few web service backends using tokens.

Work on data access tokens for interactive jobs will start in autumn of 2025. A prototype of utilities for users to acquire tokens and execution wrappers need to be developed.

CMS tokens are currently quite rich in capabilities. The plan is to refine and limit the scopes inside tokens over the coming years. This includes finer path specification for storage access, from the current top namespace to the dataset level.

In 2026, we expect to run grid jobs with only tokens. The goal of the experiment is to decouple from X.509 user certificates well before the start of HL-LHC in 2030.

## Acknowledgements

This document was prepared by CMS using the resources of the Fermi National Accelerator Laboratory (Fermilab), a U.S. Department of Energy, Office of Science, Office of High Energy Physics HEP User Facility. Fermilab is managed by Fermi Forward Discovery Group, LLC, acting under Contract No. 89243024CSC000002.